\documentclass[twocolumn,showpacs,preprintnumbers,amsmath,amssymb]{revtex4}
\usepackage{graphicx}
\usepackage{dcolumn}
\usepackage{bm}
\begin{document}
\preprint{APS/123-QED}
\title{About mechanics of the structured particles }
\author{V.M. Somsikov}
 \altaffiliation[] {}
 \email{vmsoms@rambler.ru}
\affiliation{%
Laboratory of Physics of the geoheliocosmic relation, Institute of
Ionosphere, Almaty, Kazahstan.
}%

\date{\today}
\begin{abstract}

The principles of creation of the mechanics of structured
particles in the frame of the Newton's laws are considered. The
explanation how this mechanics leads to the account of dissipative
forces is offered. It is discussed why the motions of the system
determine by two type of symmetry: symmetry of the system and
symmetry of space and how it leads to two types of energy and
forces accordingly. It is shown how the mechanics of the
structured particles leads to thermodynamics, statistical physics
and kinetics. \end{abstract}

\maketitle

\section{\label{sec:level1}Introduction\protect}

The Newton's motion equation is gained on the basis of the model
bodies in the form of the material points (ÌP) and solid bodies.
Such idealization of models of real bodies leads us to the second
law of Newton. According to this law, the acceleration of MP is
proportional to the potential force which acts on it [1, 2]. The
work of this force is equal to their integral along the way. The
energy conservation law of MP from here follows. In connection
with this law the dynamics of MP is determined by two types of
energy: the kinetic energy and potential energy. Along trajectory
of MP the sum of these types of energy is constant. The MP motion
is reversible. It is follows from the Newton's second law.

All bodies in the nature have a structure. They have the internal
energy which is caused by relative motion of the body's elements.
Therefore the works of the external forces change not only the
body's motion energy but the internal energy also. However the
Newton's motion equation, which has been constructed on the basis
of models of structureless bodies, does not include the terms
responsible for the change an internal energy. In practice the
change an internal energy are taken into account by addition to
the Newton's motion equation of the empirical force of a friction.

The work of the frictional forces defines the dissipative part of
motion energy which goes to the body's internal energy and
dissipated in the environment [2]. The friction coefficient is
taken from the experiment. Thus, the rigorous description of the
dynamics of bodies in the frame of classical mechanics is absent.
It is due to the simplification of the bodies models. We have
assumed from here that for description of a motion of real bodies,
the MP should be replaced on a structural particle and the motion
equation for the structural particles should be obtained.

The great diversity of structural particles does not allow
analyzing all types of energy dissipation. But we can select such
relatively simple models that allow understanding the nature of
dissipation in the framework of the laws of classical mechanics.
We found that it is a system of potentially interacting material
points.

The problem of description of the dissipative forces in the frame
of the classical mechanics is similar to the problem of
irreversibility. This problem was formulated by Boltzmann. All
attempts to solve it without the use of statistical laws were till
now unsuccessful. The generally accepted explanations of the
irreversibility of today are based on probabilistic laws
contradicting the determinism of classical mechanics [3].
Nevertheless, the explanation of irreversibility without
attraction of probabilistically laws, if particles possess by the
structure, can be offered [4, 5].

To find an approach to solving the irreversibility problem in the
framework of the laws of classical mechanics, the dynamics of hard
disks was studied in the beginning. As a result, it was found that
the system, consisting of two interacting of disks subsystems,
moves to equilibrium [4]. It has been shown that this is due to
the transformation of energy of relative motion of subsystems into
the motion energy of disks relative to the center of masses (CM)
of the corresponding subsystem. The same mechanism of
equilibration takes place for the structured particles (SP) where
SP is equilibrium system consisting from a big enough number of
potentially interacting MP.

Mechanic of SP can be constructed at following restrictions [6]:
1). Everyone MP is belonging to its SP during all process. 2). SP
is in equilibrium during all time. The first restriction
eliminates inessential complications related to the necessity to
reconsider of SP structure due to transitions ÌP between them. The
second restriction is equivalent to the requirement of weak
interacting which accepted in thermodynamics.

The aim of this paper is to show how the mechanics of SP can be
constructed on the bases of a Newton's laws for MP. For this
purpose the nature of the restrictions of classical mechanics is
analyzed. The explanation of the necessity of the systems dynamics
description on the basis of two types of symmetry: the symmetry of
the system and the symmetry of the space are submitted. How from
SP mechanics to come to thermodynamics, statistical physics and
kinetics and how to introduce the concept of entropy into the
classic mechanics are explained.

\section{The system of two MP}
The basic principles for construct of the SP mechanics as well as
the method of its construction can be illustrated on the example
of system of two MP. The task of two MP is solved by transition to
a coordinate system of CM [7]. In this case, the variables are
separated. The nature of such separation of variables is connected
with the emergence of a new quality of a system that is absent for
MP. It is internal energy caused by the relative motions of system
elements. The energy of two MP in laboratory coordinates of system
(LS) has the form:
\begin{eqnarray}
m({v^2}_1+{v^2}_2)/2+U(r_{12})+U^{env}(r_1)+U^{env}(r_2)=const
\label{eqn1}
\end{eqnarray}
where $U(r_{12})$ is a potential energy of MP interaction;
$U^{env}(r_1)$, $U^{env}(r_2)$ are potential energies for MP in an
external field of forces; $r_1, r_2$-coordinates of MP;
$r_{12}=(r_1-r_2)$; $v_1, v_2$ are the velocities of MP.

The motion each MP is caused by two independent types of forces:
forces of interaction MP and external forces. In the LS coordinate
system the task is nonlinear because the motion of one MP depends
from the motion of the other MP. Thus in the LS coordinates of
system the MP motion are interdependent. Therefore the LS system
is unacceptable for the description of dynamics of system. New
variables are set as follows: $R_2=(r_1+r_2)/2, V_2=
\dot{R}_{12}$, are coordinates and velocities for CM,
$v_{12}=\dot{r}_{12}$. In these variables the system's energy is:
\begin{eqnarray}
E={MV^2}/2 +mv_{12}^2/4+U(r_{12})+U^{env}(R_2,r_{12}) \label{eqn2}
\end{eqnarray}
Here ${MV^2}/2$ is a kinetic energy of CM system's motion. The
energy $mv_{12}^2/4+U(r_{12})$ is a internal energy of system
determined by forces of interaction MP and their relative motion;
; $U^{env}(R_2,r_{12})$ is a potential energy of system in an
external field. $M=2m$.

Differentiating the energy (2) with respect to time, we get:
\begin{eqnarray}
MV\dot{V}+mv_{12}\dot{v}_{12}/2+F_{12}v_{12}+
{F_{R_2}}^{env}V_2+{F_{r_{12}}}^{env}v_{12}=0 \label{eqn3}
\end{eqnarray}
where $F_{12}=\partial{U({r}_{12})}/\partial{r_{12}}$,
$F_{R_2}^{env}=\partial(U^{env})/\partial{R_2}$,
$F_{r_{12}}^{env}=\partial(U^{env}))/\partial{r_{12}}$.

If there is no external force field, the last two terms in eq. (3)
are zero. Variables are separated and eq. (3) is integrable. If
the external field exist but does not depend from $r_{12}$ then
last term in the eq. (3) is equal to zero and its breaks up on two
independent equations:
\begin{eqnarray}
MV_2\dot{V}_2+F^{env}_{R_2}V_2=D
\end{eqnarray}
\begin{eqnarray}
mv_{12}\dot{v}_{12}/2+F_{12}v_{12}=-D
\end{eqnarray}
Here the eq. (4) describes the motion of the CM system in an
external field; the eq. (5) describes the relative motion MP which
does not depend on exterior forces; $D$ is a constant which could
be chosen equal to zero. It means that when the external forces
are homogeneous the internal energy can't change. Thus in the
first and second cases the motion of two MP is determined by the
Newton's third law. In generally the exterior forces can change
both the energy of system motion and internal energy.

Thus, on the example of the two-body system has shown that the
energy of the system is split into two independent types by
transition to the CM coordinates system. It is the internal energy
which depends on the relative velocities of MP and the forces of
their interaction. And it is the energy of the system motion in
the field of external forces which depends on the coordinates of
the CM and its velocity. We can see that by summarizing of the
motion equation for LS, we exclude internal forces, leaving only
the external forces. As a result we come to the system's motion
equation in space. By subtracting these equations, we exclude the
external forces and come to the equation defining the relative
motion of the MP due to their interaction forces. I.e. the
system's motion, unlike the MP motion, is determined by two
invariants: the energy of its motion and internal energy.

All bodies consist of micro-particles or molecules. Therefore they
can be represented in the form of the SP whose position is
determined by its CM. As shown on example of two MP, the motion of
each MP should be determined in relative to the CM. Coordinates
and velocities of the MP relative to the CM systems we will call
micro variables, the coordinates and velocity of the CM systems we
will call as macro variables. Since internal and external forces
are independent, then these variables are also independent. Hence
the two spaces variables in relevant micro and macro variables,
also independent. I.e. the new variables divide the space of the
generalized co-ordinates and velocities on two independent
subspaces. One subspace is determined by the internal symmetries
of the system, and the second subspace is determined by the
symmetry of the outer space [8]. Thus the systems dynamics is
defined by two types of symmetry: symmetry of the system which
defined by distribution of its elements and character of their
interactions, and symmetry of space in which the system moves.
Hence, the energy of the system will be the sum of two invariants
of motion: internal energy and the energy of motion of the system
as a whole.

Since the energy, unlike the forces, is the additive function of
dynamic parameters of the MP, the mechanics of SP conveniently
builds basing on the energy function. Below we will obtain the
expression for the energy of the system consisting from
potentially interacting MP which will be written down in micro and
macro dynamic variables.

\section{Energy of the MP system}
Let us take a system from $N$ of potentially interacting a unit
mass MP. The potentials in each point of space are additive.
Therefore the force acting on a given MP is equal to the sum of
forces acting on it from all others MP and from the external
forces. The forces between everyone two MP are determined by
distance between them. Thus the kinetic energy of system $T_N$
can be represented as the sum of the kinetic energies of the MP.

So $T_N=\sum\limits_{i=1}^{N} m{v_i}^2/2$. Potential energy is
equal to the sum of potential energies of all MP in the field of
the external forces and potential energies of MP pair interactions
among themselves which is
${U_N}(r_{ij})={\sum\limits_{i=1}^{N-1}}{\sum\limits_{j=i+1}^{N}}U_{ij}(r_{ij})
$, where $i,j=1,2,3...N$ are the numbers of MP, $r_i, v_i$ are
coordinates and velocities of the $i$-th MP, $r_{ij}=r_i-r_j$.
Hence, full energy of system is equal to
$E=E_N+U^{env}=T_N+U_N+U^{env}=const$.

It is obvious that kinetic energy of system includes the energy of
its motion in the field of external forces, $T_N^{tr}$ and kinetic
energy of relative motion, $T_N^{ins}$ caused by interactions MP
among themselves. I.e., $T_N=T_N^{ins}+T_N^{tr}$. We will write
down the velocities of everyone MP in the form of the sum:
$v_i=V_N+\tilde{v}_i$, where $V_N=\dot{R}_N$,
$R_N=(\sum\limits_{i=1}^{N}r_i)/N$,
$\sum\limits_{i=1}^{N}\tilde{v}_i=0$. Hence
$T_N=\sum\limits_{i=1}^{N}
m{v_i}^2/2=M_NV^2_N/2+\sum\limits_{i=1}^{N}{m\tilde{v}_i^2/2}$. It
means that unlike one MP the total kinetic energy of motionless
system is equal to the sum kinetic energies of MP determined by
their velocities relative to the CM. I.e. one part of kinetic
energy of system is connected with motion of the MP relative to
the CM and the second part is connected with the motion of the
system CM. Hence, the velocity of system is determined by the
velocity of its CM whose position is defined by a radius-vector
$R_N$.

Thus the system's energy consists of the kinetic energy of the MP
motion relative to the CM and the potential energy of their
interaction. The sum of this energy called the internal energy of
the system. Then the energy can be written as a sum of internal
energy plus the system's energy in the field of external forces.
I.e.:
\begin{eqnarray}
E_N=T_N^{tr}+E_N^{ins}+U^{env}, \label{eqn6}
\end{eqnarray}
where $E_N^{ins}=T_N^{ins}+U_N$, is internal energy,
$T_N^{ins}=\sum\limits_{i=1}^{N}m\tilde{v}_i^2/2$ is a kinetic
part of internal energy, $U_N$ is a potential part of internal
energy, determined by the interactions of MP.

Quadratic function of the kinetic energy can be expressed through
a quadratic function in which arguments are the velocities of the
MP in relative to the CM and the velocity of the CM system. This
conclusion is follow from the equality: $\sum\limits_{i=1}^{N}
m{v_i}^2/2=m/(2N)\{V_N^2+\sum\limits_{i=1}^{N-1}\sum\limits_{j=i+1}^{N}v_{ij}^2\}$
(a). The first term in (a) is the kinetic energy of the CM motion.
The second term is the kinetic part of the internal energy
determined by the relative velocities of MP.

Let's transform the energy $T_N$ by replacement:
$v_i=V_N+\tilde{v}_i$, where $\tilde{v}_i$ is a MP velocities
relative to the CM. As $\sum\limits_{i=1}^{N}\tilde{v}_i=0$, then
$T_N=M_NV_N^2/2+\sum\limits_{i=1}^{N} m{\tilde{v}_i}^2/2$. Using
(a) we will find: $\sum\limits_{i=1}^{N}
m{\tilde{v}_i}^2/2=\sum\limits_{i=1}^{N-1}\sum\limits_{j=i+1}^{N}v_{ij}^2$.
Therefore $T_N^{ins}=\sum\limits_{i=1}^{N}
m{\tilde{v}_i}^2/2=\sum\limits_{i=1}^{N-1}\sum\limits_{j=i+1}^{N}v_{ij}^2$.

Thus the law of energy conservation for the system can be
formulated as follows: the sum of the system's kinetic energy of
motion, its internal energy and of the potential energy in the
external field of forces always is a constant along the trajectory
of the CM. The difference of the energy conservation laws for the
system and for MP leads to a qualitative distinction for their
motions. Indeed, the trajectory MP is defined by transformation of
potential energy of an external field only into the kinetic energy
of its motion. But the trajectory of system is defined by
transformation of potential energy of an external field both to
its kinetic energy and to internal energy. The natural variables
that define these types of energy are macro and micro variables.

\section{The system's motion equation}

There are basic differences of dynamics of the systems possessing
structure and the sizes, from dynamics of MP. The motion of MP is
uniquely determined by the point in space. But SP motion is
determined by the area of space occupied with it. Therefore for
unequivocal definition of dynamics of system it is necessary to
know, both change of its kinetic energy of the system motion and
change of internal energy as energy of an external field goes on
change of these two types of energy. The system's motion depends
from its sizes if the is exist the spatial nonhomogeneity of
external forces.

Another fundamental difference between the dynamics of MP and the
dynamics of the system base on the fact that for one MP the
principle of superposition of forces is valid, while for the
different MP it is not so. Indeed the change of the internal
energy has a place when the sum of internal forces is equal to
zero. Therefore if to summarize the equations of motion for each
MP, we will lose the terms which determine the change of the
internal energy. But the system's motion is determined by the
change of two types of energy: the system's motion energy and
internal energy. Therefore the motion equation should be obtained
from the energy equation in the variables that determine the
motion of its CM and motion of the MP relative to the CM.

Differentiating the energy of the system (6) over time we obtain [5, 6]:
\begin{eqnarray}
V_NM_N\dot{V}_N+{\dot E}_N^{ins}=-V_NF^{env}-\Phi^{env}\label{eqn7}
\end{eqnarray}
Here $F^{env}=\sum\limits_{i=1}^{N}F_i^{env}(R_N,\tilde{r}_i)$,
${\dot E}_N^{ins}={\dot T}_N^{ins}(\tilde{v}_i)+{\dot
U}_N^{ins}(\tilde{r}_i)$=
$\sum\limits_{i=1}^{N}\tilde{v}_i(m\dot{\tilde{v}}_i+F(\tilde{r})_i)$,
 $\Phi^{env}=\sum\limits_{i=1}^{N}\tilde{v}_iF_i^{env}(R_N,\tilde{r}_i)$,
$r_i=R_N+\tilde{r}_i$, $M_N=mN$, $v_i=V_N+\tilde{v}_i$,
$F_i^{env}=\partial{U^{env}}/\partial{\tilde{r}_i)}$, $\tilde{r}_i$,
$\tilde{v}_i$ are the coordinates and velocity of $i$-th particle in
the CM system, $R_N,V_N$ are the coordinates and velocity of the CM
system.

The equation (7) represents the balance of the energy of the system
of material points in the field of external forces.

The first term
in the left-hand side of the equation determines the change of
kinetic energy of the system - ${\dot{T}}_N^{tr}=V_NM_N\dot{V}_N$.
The second term determines the change of internal energy of the
system. This energy dependent on coordinates and
velocities of material points relative to the CM.

The right-hand side corresponds to the work of external
forces changing the energy of the system. The first term changes
the systems motion energy. The second term determines the
work of forces changing the internal energy.

Now let us take the external forces which scale of heterogeneity
is commensurable with the systems scales. In this case we can
write: $F^{env}=F^{env}(R+\tilde{r}_i)$ where $R$ is a distance to
the CM. Let us assume that $R>>\tilde{r}_i$. In this case the
force $F^{env}$ can be expanded with respect to a small parameter.
Leaving in the expansion terms of zero and first order we can
write:
$F_i^{env}=F_i^{env}|_{R}+(\nabla{F_i^{env}})|_{R}\tilde{r}_i\equiv
F_{i0}^{env}+(\nabla{F_{i0}^{env}})\tilde{r}_i$. Taking into
account that $\sum\limits_{i=1}^{N}\tilde{v}_i
=\sum\limits_{i=1}^{N}\tilde{r}_i=0$ and
$\sum\limits_{i=1}^{N}F_{i0}^{env}=NF_{i0}^{env}=F_0^{env}$, we
get from eq. (7):
\begin{eqnarray}
V_N(M_N\dot{V}_N)+
\sum\limits_{i=1}^{N}m\tilde{v}_i(\dot{\tilde{v}}_i+F(\tilde{r})_i)\approx\nonumber\\\approx
-V_NF_0^{env}-({\nabla}F^{env}_{i0})\sum\limits_{i=1}^{N}\tilde{v}_i\tilde{r}_i\label{eqn8}
\end{eqnarray}

The work of the potentially part of force, $F_0^{env}$, change the
system's motion energy. The term in the right-hand side has a
first order of smallness as the condition $R>>\tilde{r}_i$ does
not mean smallness of the $\tilde{v}_i$. This term is proportional
of the gradient of external force and determines the work on
change of internal energy. Its variation can't be expressed by the
integral of the gradient of scalar function on the way. It is
because the change of internal energy is a sum of work of external
forces in change the relative motion of MP. Therefore these forces
can be expressed through the effectiveness of the change in
internal energy. This can be done so. Multiplying eq.(7) by $V_N$
and dividing by $V_N^2$ we find the equation of system motion [6]:
\begin{eqnarray}
M_N\dot{V}_N= -F^{env}-{\alpha_N}V_N\label{eqn9}
\end{eqnarray}
where $\alpha_N=[{\dot E}_N^{ins}+\Phi^{env}]/V_N^2$.

The second term in the right-hand side defines a non-potential
part of forces whose work changes the internal energy. If the
external field of force is homogeneous or when the forces between
MP are much more then acting on the system of the external forces,
this term is equal to zero and the eq. (9) becomes the Newton's
motion equation.

Thus, to obtain the motion equation for the structured body, it is
necessary to execute consistently the following operations.
Firstly, it is necessary to present a body as a system of micro
particles. By transition to micro and macro parameters we present
the system's energy as a sum of the motion energy and an internal
energy. Then we obtain the equation of energy flow between these
types of energy. From here we will come to a system's motion
equation which takes into account non-potential force changing its
internal energy.

It is important to note that the eqs. (6-9) strictly follow from
Newton's laws for MP. Therefore, all properties of the systems
dynamics follow from these laws.

\section{The system of SP}

The above equations are valid for the general case of any systems
of potentially interacting MP in the external field of forces. In
generally due to the nonlinearities they are not integrable. But
integration is possible if the system represents as a set of SP.
The equilibrium of SP means that it can be split on the rather
large equilibrium subsystems which are motionless relative to each
other. The SP internal energy is the sum of the internal energies
of subsystems. I.e. the collective processes of energy, momentum
and mass flows into SP are absent. Therefore at feeble enough
action on SP not breaking equilibrium, its motion will be
determined by the change of the motion energy and an internal
energy.

In the approach of the local equilibrium approximation any
non-equilibrium system can be represented by a set of SP which has
a relative motion to each other. In the thermodynamic limit at
enough weak interactions, each of the SP during the entire process
can be regarded as equilibrium [9]. Then the dynamics of
non-equilibrium systems can be described by the eq. (9).

Let us the system consists of two SP: $L$  and $K$ . Let us $L$ is
the number of elements in the $L$-SP and $K$ is the number of
elements in $K$-SP, i.e. $L+K=N$.

Let us CM for two SP motionless, i.e. $LV_L+KV_K=0$, where $V_L$
and $V_K$ are velocities of $L$ and $K$ equilibrium subsystems
relative to the CM system. Differentiating the energy of the
system with respect to time, we obtain:
${\sum\limits_{i=1}^{N}v_i{\dot{v}}_i}+{\sum\limits_{i=1}^{N-1}}\sum\limits_{j=i+1}^{N}v_{ij}
F_{ij}=0$, where $F_{ij}=U_{ij}=\partial{U}/\partial{r_{ij}}$. In
order to derive the equation for $L$-SP, in the left-hand side of
the equation we leave only terms determining change of kinetic and
potential energy of interaction of $L$-SP elements among
themselves. All other terms we displace into the right-hand side
of the equation and combine the groups of terms in such a way that
each group contains the terms with identical velocities. In
accordance with Newton equation, the groups which contain terms
with velocities of the elements from $K$-SP are equal to zero. As
a result the right-hand side of the equation will contain only the
terms which determine the interaction of the elements $L$-SP with
the elements $K$-SP. Thus we will have:
${\sum\limits_{i_L=1}^{L}}v_{i_L}
{\dot{v}}_{i_L}+{\sum\limits_{i_L=1}^{L-1}}\sum\limits_{j_L=i_L+1}^{L}
F_{{i_L}{j_L}}v_{{i_L}{j_L}}={\sum\limits_{i_L=1}^{L}}\sum\limits_{j_K=1}^{K}
F_{{i_L}{j_K}}v_{j_K}$ where double indexes are introduced to
denote that a particle belongs to the corresponding system. If we
make substitution $v_{i_L}=\tilde{v}_{i_L}+V_L$, where
$\tilde{v}_{i_L}$ is the velocity of $i_L$ particle relative to
the CM of $L$ -SP, we obtain the equation for $L$-SP. The equation
for $K$-SP can be obtained in the same way. The equations for two
interacting systems can be written as [6, 11]:
\begin{eqnarray}
V_LM_L\dot{V}_L+{\dot{E}_L}^{ins}=-{\Phi}_L-V_L{\Psi}
\end{eqnarray}
\begin{eqnarray}
V_KM_K\dot{V}_K+{\dot{E}_K}^{ins}={\Phi}_K+V_K{\Psi}
\end{eqnarray}

Here $M_L=mL, M_K=mK, \Psi=\sum\limits_{{i_L}=1}^LF^K_{i_L}$;
${\Phi}_L=\sum\limits_{{i_L}=1}^L\tilde{v}_{i_L}F^K_{i_L}$;
${\Phi}_K=\sum\limits_{{i_K}=1}^K\tilde{v}_{i_K}F^L_{i_K}$;
$F^K_{i_L}=\sum\limits_{{j_K}=1}^KF_{i_Lj_K}$;
$F^L_{j_K}=\sum\limits_{{i_L}=1}^LF_{i_Lj_K}$;
${\dot{E}_L}^{ins}={\sum\limits_{i_L=1}^{L-1}}\sum\limits_{j_L=i_L+1}^{L}v_{i_Lj_L}
[\frac{{m\dot{v}}_{i_Lj_L}}{L}+\nonumber\\+F_{i_Lj_L}]$;
${\dot{E}_K}^{ins}={\sum\limits_{i_K=1}^{K-1}}\sum\limits_{j_K=i_K+1}^{K}v_{i_Kj_K}
[\frac{{m\dot{v}}_{i_Kj_K}}{K}+\nonumber\\+F_{i_Kj_K}]$.

The equations (10, 11) are equations for interactions between SP.
They describe energy exchange between SP. Independent variables
are macro-parameters and micro-parameters. Macro-parameters are
coordinates and velocities of the motion of CM of systems.
Micro-parameters are relative coordinates and velocities of
material points. Therefore the equation of system interaction
binds together two types of description: on the macro-level and on
the micro-level. The description on the macro-level determines
dynamics of an SP as a whole and description on the micro-level
determines dynamics of the elements of an SP.

The potential force, $\Psi$, determines the motion of an SP as a
whole. This force is the sum of potential forces acting on the
elements of one SP from the other system.

The forces determined by terms ${\Phi}_L$ and ${\Phi}_K$ transform
the motion energy of SP into their internal energy as a result of
chaotic motion of elements of one SP in the field of forces of the
other SP. As in the case of the system in the external field,
these terms are not zero only if the characteristic scale of
inhomogeneity of forces of one system is commeasurable with the
scale of the other system. The work of such forces causes
violation of time symmetry for SP dynamics.

The equations for systems motion corresponding to the equations
(10,11) can be written as:
\begin{equation}
M_L\dot{V}_L=-\Psi-{\alpha}_LV_L \label{eqn12}
\end{equation}
\begin{equation}
M_K\dot{V}_K=\Psi+{\alpha}_KV_K\label{eqn13}
\end{equation}
where ${\alpha}_{L}=(\dot{E}^{ins}_{L}+{\Phi}_{L})/V^2_{L}$,
${\alpha}_{K}=({\Phi}_{K}-\dot{E}^{ins}_{K})/V^2_{K}$,

The eqs. (12, 13) are written down for SP which are considered
equilibrium during all process of interaction. In this case we can
neglect by the energy, momentum and mass flows into SP. Due to
equilibrium of SP its internal energy can't be transformed into SP
motion energy. This follows from the law of conservation of
momentum, according to which neither any internal MPs motions can
change of SP velocity. From here we come to a conclusion about
irreversibility of SP dynamics. Therefore we can call the
"$\alpha_L$", "$\alpha_K$"  as a friction coefficients.

Dynamics of non-equilibrium systems are determined by the eqs.
(12, 13). Consequently the Lagrange Hamilton and Liouville
equations for the systems whose elements are the SP will also be
determined by these equations. It is well known that the Hamilton
principle for MP derived from differential D'Alambert principle
using Newton equation for MP [2]. For this purpose the time
integral of virtual work $\delta\omega^e$ done by effective forces
is equated to zero. Integration over time is carried out provided
that external forces possess a power function. It means that the
canonical principle of Hamilton is valid only for cases when $\sum
F_i\delta R_i=-\delta U$, (b) where $i$ is a particle number, and
$F_i$ is a force acting on this particle. But for interacting SP
the condition of conservation of forces is not fulfilled because
of the presence of a non-potential component. Therefore in the
equations of Lagrange, Hamilton and Liouville for systems from SP
the terms caused by non-potentiality of collective forces are
appeared. In this case the Liouville equation looks like [4, 6]:
\begin{equation}
df/dt=-\sum\limits_{L=1}^{R}{\partial}{F_L}/{\partial}V_L
\label{eqn14}
\end{equation}

Here $f$ is a distribution function for a set of SP, $F_L$ is a
non-potential part of collective forces acting on the SP, $V_L$ is
the velocity of $L$-SP.

The right-hand side of the equation is determined by the efficiency
of transformation of the SP motion energy into their internal
energy. For non-equilibrium systems the right-hand side is not equal
to zero because of non-potentiality of forces changing the internal
energy.

The state of this system can be defined in the phase space which
consists of   $6R-1$ coordinates and momentums of SP, where $R$ is
a number of SP. Location of each SP is given by three coordinates
and their moments. Let us call this space us $S$-space for SP in
order to distinguish it from the usual phase space for MP. It is
caused by transformation of the motion energy of SP into their
internal energy. The SP internal energy can't be transformed into
the SP energy of motion as SP momentum can't change due to the
motion of its MP. Therefore $S$-space is compressible because the
internal energy will increase until the relative motion of SP did
not disappear.

In connection with $S$-space it is necessary to redefine the
geometrical concept of an interval [2] for systems whose elements
are the SP. Indeed, we have shown that the dynamics of the SP is
determined by two types of symmetry: the internal symmetry and the
symmetry of the space. Therefore the motion of the system is
determined by two types of energy: kinetic energy of the SP and
its internal energy. Each of these types of energy has its own
type of forces. This is reflected in the fact that the geometry of
motion of the SP, in contrast to the geometry of motion of the MP,
is the sum of the squares of the two intervals, that can be
written as [2, 7]: $d\bar{s}^2=ds^2_{tr}+ds^2_{ins}$. Here
$ds^2_{tr}$ is a square of an interval corresponding to SP motion
energy, $ds^2_{ins}$ - is a square of the interval corresponding
to SP internal energy.

Thus, the square of the interval of a non-equilibrium system
splits into the sum of the squares of the two intervals. The first
one corresponds to the system motion while the second corresponds
to its internal energy. These intervals are orthogonal since they
satisfy the Pythagorean Theorem.

\section{Mechanics of SP and thermodynamics}

Difficulties of substantiation of the empirical laws of
thermodynamics based on fundamental laws of physics are connected
with the reversibility of the Newton's motion equations. The
reversibility is due to its constructing on the bases of
unstructured body's models. Acceptance in attention of structure
leads to occurrence of non-potential component of the collective
forces of body's interaction changing their internal energy. For
SP this energy can only increase due to the energy of its motion.
It is equivalent to irreversibility of the SP dynamics. Let us
explain this conclusion.

The presence of reversible dynamics for SP would mean that its
internal energy is capable to pass into motion energy. In turn
this would mean the possibility of increasing momentum of SP at
the expense of its internal energy. But this contradicts the law
of conservation of momentum. Indeed, for each of the equilibrium
subsystems into which splits SP the sum of the velocities of MP in
subsystems and sum of their interaction forces are equal to zero.
But for appearing of SP momentum it is necessary that at least in
one of subsystems the requirement of equality to zero of the sum
of forces has been disrupted. It is impossible because according
to a law of momentum conservation any of subsystems cannot acquire
a relative velocity or due to internal MP motion or forces from
unmovable subsystems. I.e. internal energy of the SP can't
transform into the energy of its motion. It is equivalent to
irreversibility. From the mathematical point of view this
conclusion follows from the fact that micro parameters determining
the MP motion are not dependent on the macro parameters that
determine the SP motion energy. This mechanism of irreversibility
is deterministic because its follows from the Newton's laws. There
is a fundamental difference between deterministic and
probabilistic mechanisms. For deterministic mechanism the
"coarse-grain" hypothesis isn't required.

Let us explain how can connect the mechanics SP and thermodynamics
[5, 6]. In thermodynamics the work of external forces breaks up on
two parts. One part is related to the reversible work. Another
part of energy goes into heat-ing system. According to it the
basic equation of thermodynamics looks like: ${dE=dQ-PdY}$. Here
$E$ is the energy of a system; $Q$ is the thermal energy;   $P$ is
the pressure; $Y$ is the volume. As we deal with equilibrium
systems, then $dQ=TdS$, where $T$ - temperature, $S$ - entropy.
According to the eq. (7), coming into the system energy can be
divided on two part. There are energy of relative motion of the SP
and its internal energy. It was showed [5] that in thermodynamics
to the change of the SP energy of relative motion there
corresponds the value of $PdY$, and to change of SPs internal
energy there corresponds value, $TdS$.

Let us take a motionless non-equilibrium system consisting of
"$R$" SP. Each SP consists of a great number of elements $N_L>>1$,
where $L=1,2,3...R, N=\sum\limits_{L=1}^{R}N_L$. Then the share of
energy, which goes on internal energy increasing, is determined by
the expression [5, 6]:
\begin{equation}
{{\Delta{S}}={\sum\limits_{L=1}^R{\{{N_L}
\sum\limits_{k=1}^{N_L}\int[{\sum\limits_s{{F^{L}_{ks}}v_k}/{E^{L}}]{dt}}\}}}}\label{eqn15}
\end{equation}

Here ${E^{L}}$ is the kinetic energy of $L$-SP; $N_L$ is the
number of elements in $L$-SP; $L=1,2,3...R$; ${R}$ is the number
of SP; ${s}$ is the number of external elements which interact
with ${k}$ element belonging to the $L$-SP; ${F_{ks}^{L}}$ is the
force acting on the $k$-element; $v_k$ is the velocity of the $k$-
element.

The eq. (15) can be viewed as entropy definition. Such definition
of entropy corresponds to Clausius entropy definition [9, 10].
Difference consists only that this entropy follows from analytical
expression for the change of an internal energy obtained by us on
the basis of Newton's laws. From the eq. (15), it is possible to
obtain the value of the entropy production and obtain the
conditions which necessary for sustain the non-equilibrium system
in the stationary state [6]. Thus, we will come to the basic
thermodynamic equation if in the equation (7) to carry out
standard transition to thermodynamic parameters [5, 9, 10].

Mechanics of SP leads to statistical physics and kinetics. Indeed,
the velocities of SP are determined by average values of
velocities of MP. The sum of the MP velocities relative to the CM
is equal to zero. Thus the internal energy is equivalent to the
square of fluctuation of the MP velocities relative to the
system's velocity. This means that the dynamics of the SP is
expressed through the first and second moments of the motion [9].

\section{Conclusion}

The key idea of expansion of the Newtonian mechanics allowing to
include the dissipative forces into description, consists in
replacements of MP on SP. External simplicity of this idea does
not mean its obviousness. Indeed the dynamical characteristics of
the system do not follow directly from simple plurality of
dynamical characteristics of elements. This is evident from the
fact that the structure of the system determines not only its
motion but also the collective forces of interactions.

In connection with the construction of the mechanics of the SP
requires knowledge of the principles of synthesis of the
properties of systems based on the properties of its elements
[11]. It is necessary to solving the first question: how to find
the SP motion equation on the basis of Newton's laws without
attracted of some statistical hypotheses.

It became clear as a result of studying of dynamics of two MP
systems that SP mechanics must to be built in space of micro and
macro variables. In these variables the energy of SP breaks up on
the energy of its motion and an internal energy. The SP motion
energy is expressed through macro-parameters. There are
co-ordinates and velocities of CM. Its change is connected with
the work of the external force acted on the CM of SP. The internal
energy is expressed through micro-parameters. The increasing of
internal energy is provided by the work of the external forces
which change the relative motion of MP. The internal and external
forces are independent. Therefore the SP motion energy and
internal energy are independent also. Independence internal and
external forces tell us about presence two types of symmetry. It
is symmetry of space and symmetry of system.

The major factor causing difference of SP dynamics from MP
dynamics is a structure and an internal energy. Taking SP as a
system's elements we, thereby, have supplied this elements with a
new properties - structure and an internal energy. The change of
an internal energy provided by the work of collective forces is a
cause's difference of SP dynamics from dynamics of MP.

Newton's laws were obtained for models structureless bodies. To
use these laws for determine the motion equation for real bodies
with the structure; we took a model of SP, consisting of
potentially interacting MP. Using the Newton's law for MP, we find
the motion equation for SP, taking into account changes in its
internal energy. It was done by using the expression for the
energy of the SP by transition to the variables that characterize
its dynamics. Derivation of the SP motion equation is carried out
so. We write the SP energy through independent macro and micro
variables. In these variables, it splits into SP's energy of
motion and the internal energy. Differentiating this energy in
time, we obtain the equation for the flux of the motion energy and
internal energy. From here we come to the SP's motion equation.
Dissipative forces are defined through the relation of this work
to the SP's motion energy.

Irreversibility is a new property of the SP dynamics. The
mechanism of irreversibility is related to the transformation of
SP motion energy into the internal energy and the inability of the
inverse transformation due to momentum conservation law. Because
the SP motion equation obtained on the basis of Newton's laws, the
irreversibility of the dynamics of the SP is deterministic. If we
neglect the change in internal energy, the motion of the SP will
be deter-mined by Newton's motion equation.

Irreversibility is a new property of the SP dynamics. The
mechanism of irreversibility is related to the transformation of
SP motion energy into the internal energy and the inability of the
inverse transformation due to momentum conservation law. Because
the SP motion equation obtained on the basis of Newton's laws, the
irreversibility of the dynamics of the SP is deterministic. If we
neglect the change in internal energy, the motion of the SP will
be determined by Newton's motion equation.

There are both similarities and differences between accepted today
a probabilistic explanations of irreversibility [3] and our
explanations. In the basis of probabilistic mechanism of
irreversibility is a fact of randomization of trajectories of
Hamiltonian systems in phase space due to the exponential
instability and the hypothesis of "coarse-grain" of the phase
space. In the deterministic mechanism of irreversibility both the
exponentially instability and mixing in phase space determine the
efficiency of transformation of the motion energy into the
internal energy. But the irreversibility follows from the momentum
conservation law and the non-potentiality of the forces which
transform the energy of motion into the internal energy. Therefore
the hypothesis about "coarse-grain" of the phase space for
deterministic irreversibility is not required.

In accordance with a deterministic mechanism of irreversibility in
classical mechanics the concept of entropy is appeared. This
entropy corresponds to the empirical entropy offered by Clausius
and is consistent with the mathematical form of its probabilistic
definition proposed by Boltzmann.

The SP motion equation states impossibility of existence of
structureless particles in a framework of the classical mechanics,
which is equivalent to infinite divisibility of matter.

\smallskip

\end{document}